\documentclass[]{revtex4}
\usepackage{textcomp}
\usepackage{graphicx}% Include figure files
\usepackage{bm}% bold math
\begin{document}

%\preprint{}

\title{Second harmonic generation in multilayer graphene induced by direct electric current}


\author{Anton Y. Bykov}
\email{lancaster@shg.ru}
\affiliation{Department of Physics, Moscow State University, 119991 Moscow, Russia}

\author{Tatiana V. Murzina}
\affiliation{Department of Physics, Moscow State University, 119991 Moscow, Russia}

\author{Maxim G. Rybin}
\affiliation{A.M. Prokhorov General Physics Institute, RAS, 38 Vavilov street, 119991, Moscow, Russia}

\author{Elena D. Obraztsova}
\affiliation{A.M. Prokhorov General Physics Institute, RAS, 38 Vavilov street, 119991, Moscow, Russia}


\date{\today}

\begin{abstract}
Optical second harmonic generation (SHG) is studied from multilayer graphene films in the presence of DC electric current flowing in the sample plane. Graphene layers are manufactured by chemical vapour deposition (CVD) technique and deposited on an oxidised Si(001) substrate. SHG intensity from graphene layer is found to be negligible in the absence of the DC current, while it increases dramatically with the application of the  electric current. The current-induced change of the SHG intensity rises linearly with the current amplitude and changes its sign under the reversal of the current direction to the opposite. The observed effect is explained in terms of the interference of second harmonic radiation reflected from the Si surface and that induced by the DC current in multilayer graphene.
\end{abstract}
%\pacs{78.67.Wj, 42.65.Ky, 78.47.jh}
%\keywords{}
\maketitle
%%%%%%%%%%%%%%%%%%%%%%%%%%%%%%%%%%%%%%%%%%%%%%%%%%%%%%%%%%%%%%%%%%%%%
Since its first experimental realisation in 2004 graphene continues to attract enhanced interest as a prospective material for both fundamental and applied science. Fascinating electronic properties which include electric field-effect\cite{Science666}, "chiral" quantum Hall effects \cite{Novoselov_Nature_438_197,Novoselov_Nature_Phys_2_177}, prospects for spintronics \cite{SpinNature2007} and valeytronics \cite{ValeyNatPhys2007} immediately pushed graphene research to the cutting edge of modern nanomaterial science and technology. Among the numerous problems currently being studied for graphene is the possible connection between the electron transport and the nonlinear-optical response. The importance of this task is dictated not only by needs of the applied research as allows distant probing of the electron flow in graphene devices but, perhaps, more importantly as a route to gain new comprehensive insight into its fundamental electronic properties. 

Second harmonic generation (SHG) is among the most ubiquitous methods used for probing surfaces and interfaces of centrosymmetric materials.\cite{shennature} High sensitivity to the surface and thin film properties arises from SHG being prohibited in the electric dipole approximation in the volume of a centrosymmetric medium. As a result it is generated basically at surfaces and interfaces where the central symmetry is broken. Moreover one can break the inversion symmetry by an external influence such as electric and magnetic field causing so-called field induced second harmonic generation.\cite{fed,fed1,NomokePan}

It has been demonstrated recently both theoretically and experimentally\cite{Vovochka_Eng, Khurgin} that DC electric current flowing in the plane of a centrosymetric semiconductor can break the symmetry of the electron density distribution, resulting in current-induced SHG (CSHG) which can overwhelm conventional electric-field-induced mechanism if the conductivity of the probed material is sufficiently high. Moreover, theoretical predictions\cite{Khurgin} made almost a decade before the advent of graphene demonstrate the possibility of the SHG  enhancement by 1$\sim$2 orders of magnitude in case of ballistic electron transport and in case of two-dimensional nature of the investigated electron system. 
In this paper we report the first investigation of current-induced second harmonic generation in multilayer graphene under ambient conditions.

Our specific experimental conditions are the following. Either $p,s$ -polarised output of a femtosecond Ti:Sapphire laser system operating in the wavelength range of 730$\sim$830 nm, pulse duration 100 fs, at 80 MHz repetition rate is focused into a 50 $\mu m$ size spot on the sample at 45$^{\circ}$ angle of incidence  with the peak intensity of approximately $0,8 GW/cm^{2}$. SHG radiation reflected in the direction of specular reflection was spectrally selected by BG39 color Schott filters and detected by a PMT. Sample holder could be rotated around the axis orthogonal to its surface thus allowing the azimuthal SHG studies. Application of the DC electric current was performed by mechanical adjusting of two spring-assisted Pt needles to the graphene film  distanced by 2$\div$3 mm. To ensure the the second order nonlinearity is the source of the signal at the SH frequency the conventional quadratic dependendence of the measured signal on the fundamental radiation intensity was checked.
	
Multilayer graphene film of 4$\sim$5 monolayers thickness was manufactured by chemical vapour deposition  technique on a thin polycrystalline $Ni$ foil and transferred to $SiO_{2}/Si(001)$ substrate with a 300 nm oxide layer\cite{CVD_NanoLett, CVD_Nature, Obraztsova1}. The size of the graphene-coated area was 7x7 mm$^2$ approximately. The size of single crystal domains in the film was shown to be in the range of 3-5 $\mu$m as evaluated from the SEM data and that is typical for this method of graphene composition \cite{CVD_Nature}. Such graphene films possess high electrical conductivity and is capable of preserving linear electron dispersion law \cite{PhysRevLett.99.256802, PhysRevB.73.245426, PhysRevLett.101.157601} due to stacking disorder.

\begin{figure}[t]
\centering
\includegraphics[width=8.6cm]{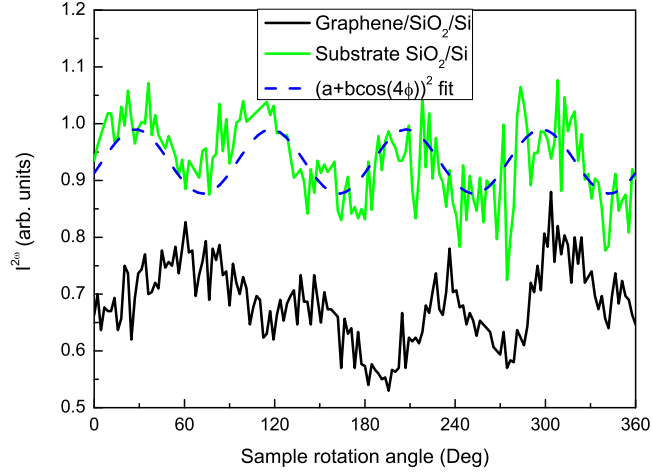}
\caption{(Color online) Azimuthal dependencies of the SHG intensity for bare Si(001) (green (light gray) line) and for graphene-coated Si(001) (black line); Fourfold sine fit (blue (dark gray) dashed line). DC electric current $J$=0}
\label{anisotropy1}
\end{figure}

As SHG is sensitive to the symmetry of the probed medium it is crucial to define the symmetry of the observed effect. Figure \ref{anisotropy1} shows a typical SHG azimuthal pattern measured in order to define the graphene film crystal anisotropy. Fourfold symmetry of the SHG intensity pattern is observed that is governed by the SHG anisotropy from Si(001), while graphene film with the structural symmetry $6mm$ should reveal three- or sixfold symmetric SHG pattern\cite{PhysRevB.82.125411}. Thus we have to consider that the main nonlinear-optical source of the sample is the Si(001) surface.
%We suggest the film lacking structural symmetry because of either layer misorientation and/or polycrystallity on laser focal spot length scale both suggested by literature\cite{CVD_NanoLett}.

\begin{figure}[Ht]
\centering
\includegraphics[width=8.6cm]{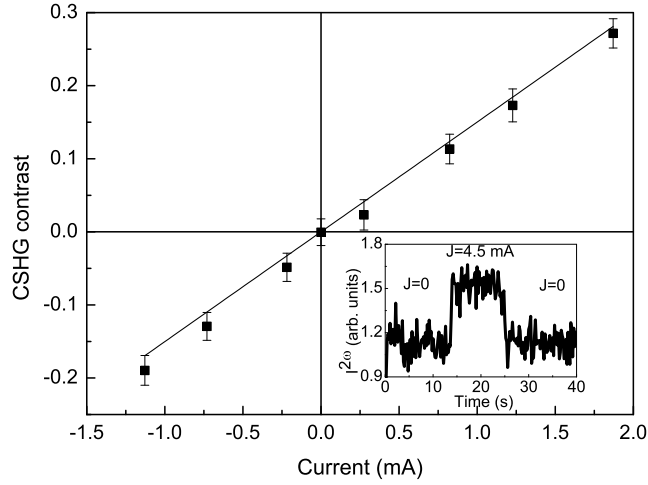}
\caption{Current dependence of the SHG contrast obtained for $pp$ combination of polarisations of the fundamental and SHG waves for $\lambda=$800 nm. Inset: Time dependence of the SHG intensity, CSHG contrast for inset $\approx$ 0.70}
\label{dependance1}
\end{figure}

A steep rise in the SHG intensity is observed as the DC current is applied (Fig. \ref{dependance1}, inset). To characterise the relative magnitude of the effect we used current-induced SHG contrast, $\rho = 2(I^{2\omega}(\vec J)-I^{2\omega}(0))/I^{2\omega}(0)$, where $I^{2\omega}(0)$ and $I^{2\omega}(\vec J)$ are the SHG intensity measured in the absence and in the presence of the current, $\vec J$. Figure \ref{dependance1} shows that $\rho(\vec J)$ dependence is linear within the experimental accuracy. This effect is consistent with the model SHG description based on the interference of the two SHG fields, $I^{2\omega} \propto (E_{0}^{2\omega}+E^{2\omega}(\vec J))^{2} \approx const+2E^{2\omega}_{Graphene}(\vec J)E^{2\omega}_{Si}Cos(\Delta\phi) \propto J$, where $E^{2\omega}_{Graphene}$ and $E^{2\omega}_{Si}$ are electric fields of second harmonic waves from graphene ans silicon, respectively, and $\Delta \phi$ is the phase shift between these fields, $J$ is the current value. Here we assume that current-induced SHG is much lower as compared with that generated by Si(001) surface.

The symmetry of the CSHG can be estimated for a material with any given crystal symmetry without even specifying the microscopic mechanism that underlines its inception. CSHG electric-dipole polarisation is given by $P_{i}^{2\omega}=\chi_{ijkl}^{(2,1)}E_{j}^{\omega}E_{k}^{\omega}J_{l}$, where $\vec E$ is the fundamental electric field and $\chi_{ijkl}^{(2,1)}$ is an effective four-rank susceptibility tensor which governs the SHG effect. 
%Taking its components for the medium with specific symmetry from \cite{shen} one after simple calculation arrives to the symmetry of the effect. 
It can be easily shown (see appendix) that the dependencies of the CSHG intensity on the angle $\phi$ between the direction of the current flow and the plane of incidence are $I_{pp,sp}\propto|a\cos\phi|^2,$ and  $I_{ps,ss}\propto|b\sin\phi|^2$ for $pp,sp$ and $ps,ss$ combinations of the input and SHG polarizations, respectively. Here 
$a,b$ are the combinations of the $\hat{\chi}^{(2,1)}$ components. 
Thus the CSHG contrast should depend on the polarization of the pump and SHG beams.

\begin{figure*}[Ht]
\centering
\begin{minipage}{.49\textwidth}
\textbf{(a)} \\
\includegraphics[width=7cm]{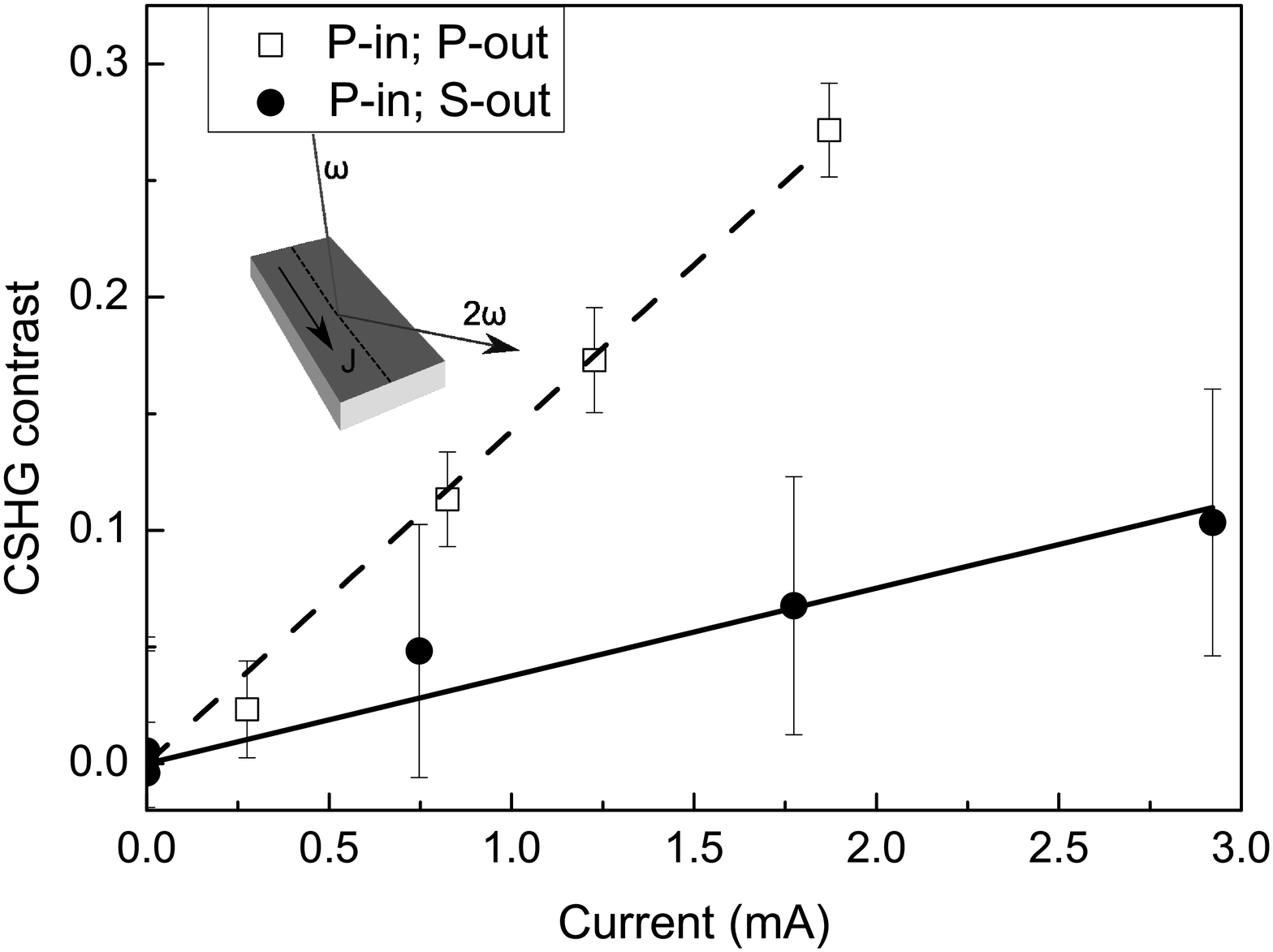}
\end{minipage}
\hfill
\begin{minipage}{.49\textwidth}
\textbf{(b)} \\
\includegraphics[width=7cm]{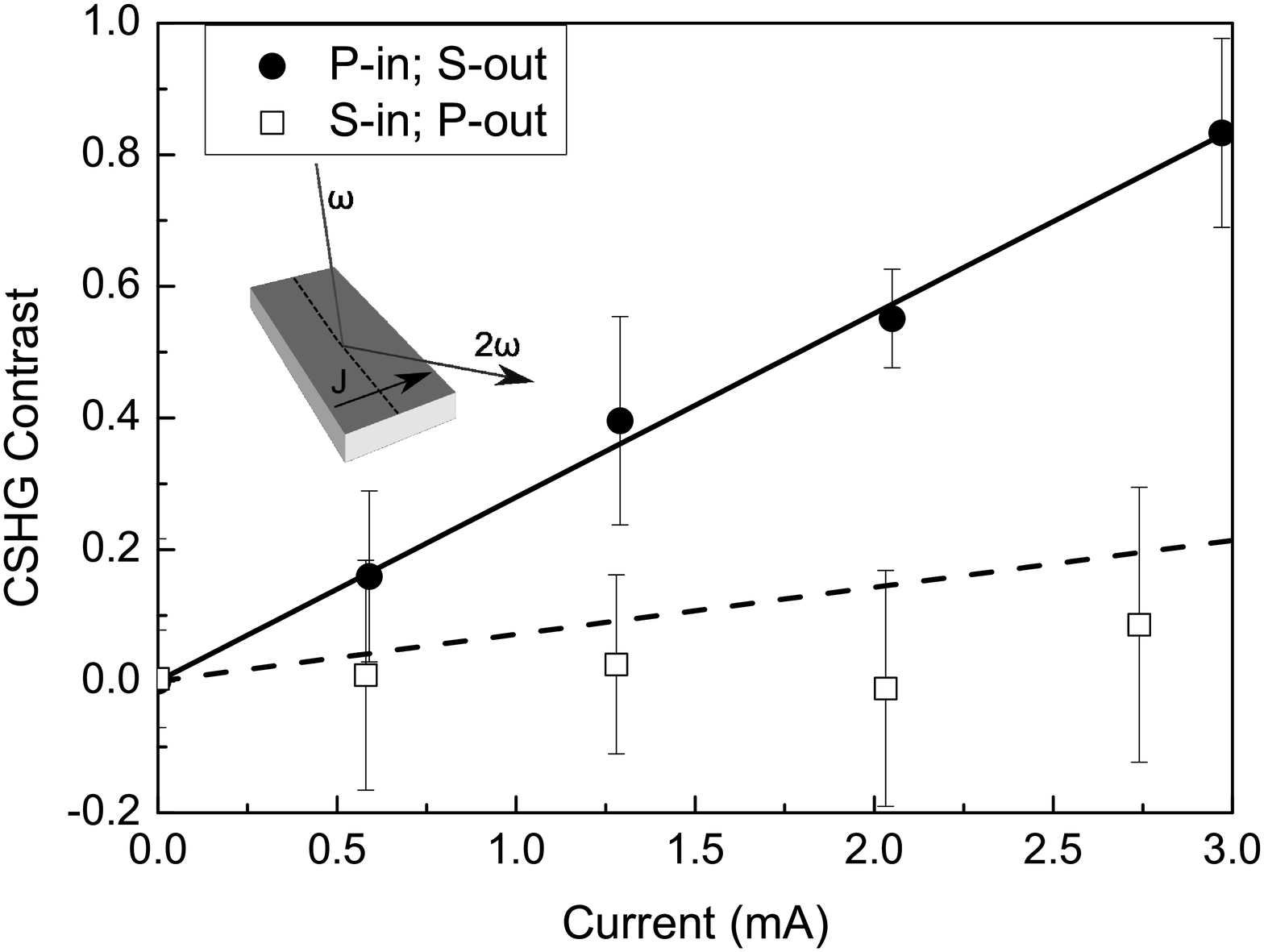}
\end{minipage}
\caption{CSHG contrast dependencies on the current value for different combinations of polarisations of the pump and SH waves ($\lambda$=800 nm)  for \textbf{(a)} $\phi$=0$^{\circ}$ and \textbf{(b)} $\phi$=90$^{\circ}$; Lines are linear fit of the experimental data}
\label{symmetry1}
\end{figure*}

Figure \ref{symmetry1} shows the dependencies of the CSHG contrast on the current for different $\phi$ values. It can be seen that in case of the current flow being parallel to the plane of incidence ($\phi$=0$^{\circ}$) the CSHG contrast for \textit{p-}polarized SHG is much higher as compared with \textit{s-}polarized one, while for $\phi$=90$^{\circ}$ the situation is inverted. This stays in agreement with the symmetry description discussed briefly above.
   
\begin{figure*}[ht]
\centering
\begin{minipage}{0.49\textwidth}
\textbf{(a)} \\
\includegraphics[width=7cm]{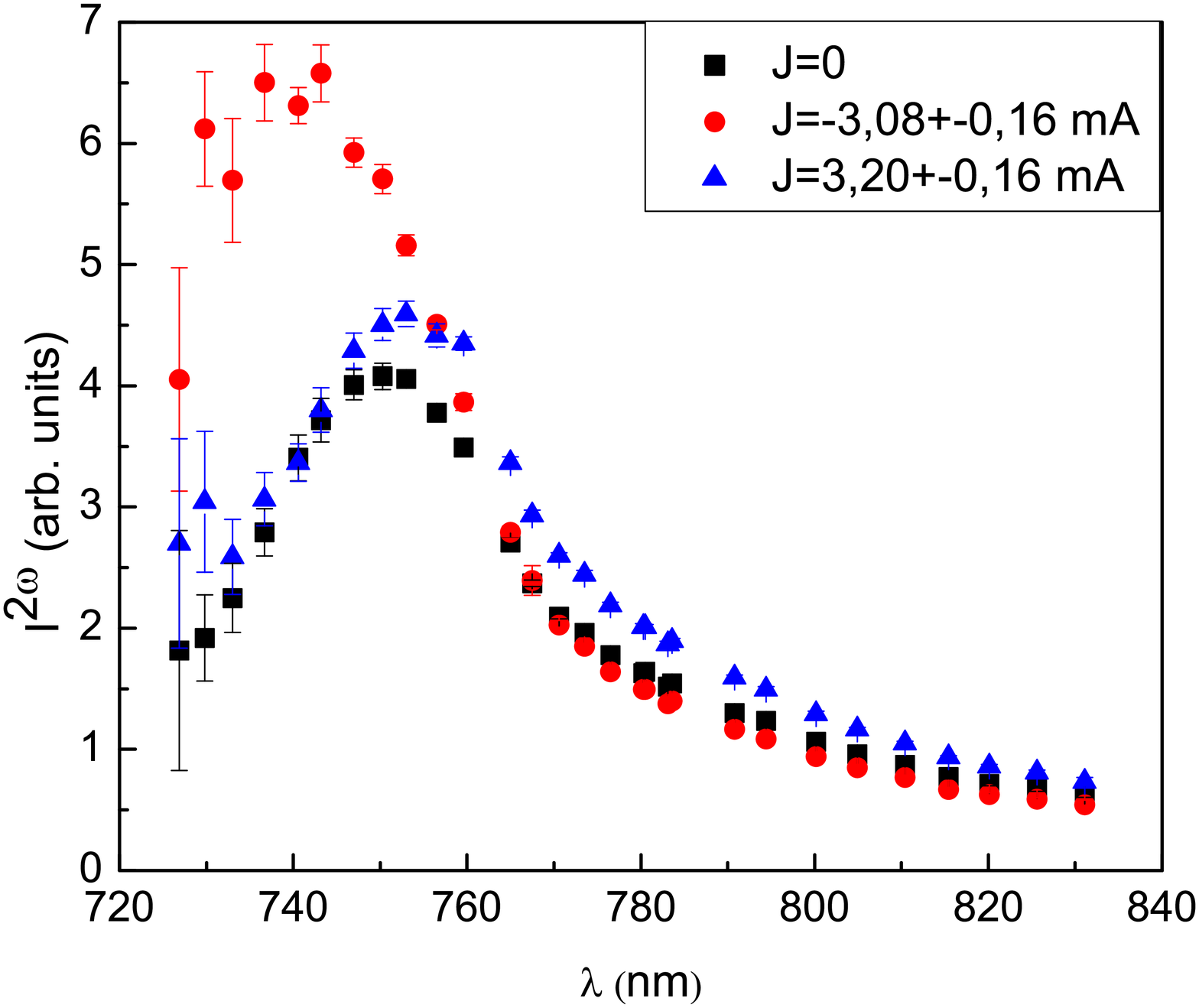}
\end{minipage}
\hfill
\begin{minipage}{.49\textwidth}
\textbf{(b)} \\
\includegraphics[width=7cm]{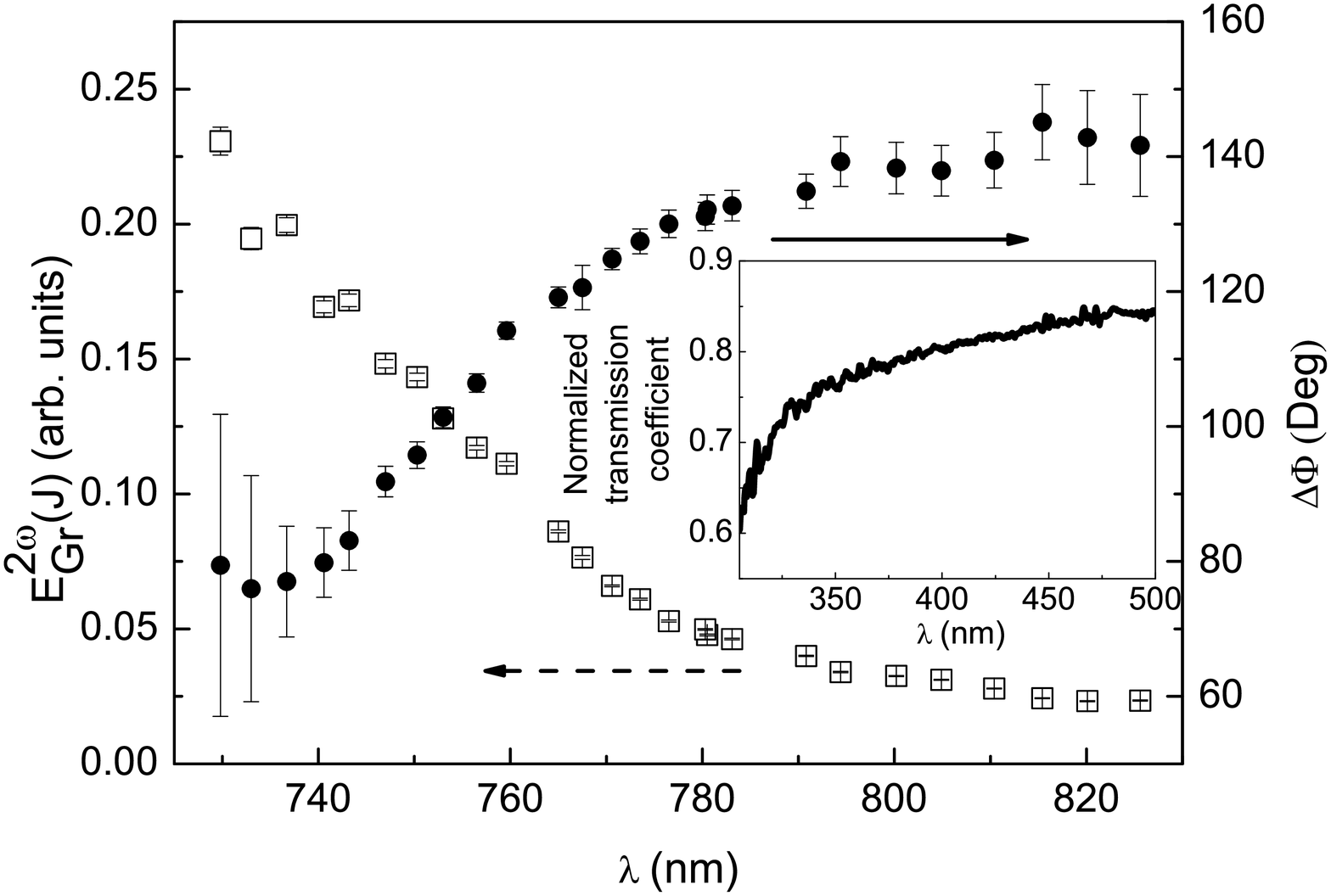}
\end{minipage}
\caption{(Color online) \textbf{(a)} SHG intensity spectrum: Black dots - SHG intensity spectrum without current, red and blue dots - SHG intensity spectrum with current in different directions \textbf{(b)} CSHG electric field (open squares) and phase shift (closed circles) spectra; Inset: Transmission spectra of the mutilayer graphene  on glass in the near-UV}
\label{spectrum1}
\end{figure*}

Finally we performed the SHG spectroscopy measurements. Figure \ref{spectrum1} shows the SHG intensity spectra obtained in the absence of the DC current as well as its current-induced modification. In the former case the SHG spectrum is consistent with that of a Si(001) surface. It demonstrates a maximum centered at approximately 3,34 eV ($\lambda \approx$ 740 nm) which corresponds to the two-photon direct interband transition in silicon\cite{fed2}. The SHG spectrum shifts significantly under the application of the DC current with the sign of this shift depending on the direction of the current .

These  spectral dependencies can be understood considering a phase shift between the interfering SH signals from the silicon surface and CSHG from graphene which varies with the wavelength. To prove this explanation right we have calculated the CSHG electric field and phase shift spectra from fig. \ref{spectrum1}a assuming that the phase of the current-induced SHG wave changes by $\pi$ with the change of the direction of the current to the opposite. The phase shift (Fig. \ref{spectrum1}b, circles) resembles conventional spectral shape in the vicinity of the resonance in silicon similar to what we have expected. The more remarkable however is the pure CSHG spectrum (Fig. \ref{spectrum1}b, open squares) which shows an enhancement at the short wavelength edge correlating with the rise of the absorption at the corresponding two-photon energy (Fig. \ref{spectrum1}b, inset). This is generally believed to come from high energy excitonic resonances at 4,5 eV \cite{weber:091904, PhysRevLett.103.186802} and/or interband transitions at 5,1 eV\cite{PhysRev.71.622}. Thus we believe the same mechanisms to play their role in the CSHG enhancement in graphene. Still with no theory proposed we cannot specify the microscopic mechanism of such enhancement and this subject seeks further investigation.

It is also necessary to mention that the observed effect may be the result of electric-field induced SHG (EFISH) driven by planar field associated with the current flow as a remarkably high $\chi^{(3)}$ has been recently reported for graphene\cite{PhysRevLett.105.097401}. However we believe that CSHG should overwhelm the EFISH in graphene because of high conductivity(resulting in insufficiently high electric field).  

In conclusion we have measured second harmonic generation from multilayer CVD graphene subjected to electron flowing within the sample plane. As the film itself was found to produce no significant SH response the situation changed drastically with in plane current application. The dependence of the CSHG effect magnitude on current density is found to be linear in a good agreement with the theory and the effect was observed to depend on current directions as was expected from the symmetry analysis. These two results combined provide a clear evidence of it being possible to use the discussed technique for distant probing of current density distribution in graphene devices. Finally a significant interference-mediated SHG spectral shift associated with current flowing in the sample and the CSHG enhancement at the short wavelength spectral edge possibly connected to resonances in graphene in UV are observed.

\begin{acknowledgements}
 The authors acknowledge helpful discussions with Prof. Oleg A. Aktsipterov and the support from the Russian Foundation of Basic Research (RFBR grants \textnumero$ $ 11-02-92121 and \textnumero$ $ 12-02-00792).
\end{acknowledgements}

\appendix*
\section{CSHG symmetry analysis}
The symmetry of the CSHG effect it determined by the symmetry of the effective forth rank susceptibility tensor $\chi^{(2,1)}$ that governs the current-induced component of the SHG field in the electric-dipole approximation, $E^{2\omega}(J) \propto \vec{P}^{2\omega}(J)=\hat{\chi}^{(2,1)} \vec{E}^{\omega} \vec{E}^{\omega}\vec{J}$. While an ideal graphene layer possesses \textit{3m} symmetry, it stems from our SHG azimuthal studies that no anisotropic SHG comes from the graphene layer in the absence of DC current. This may be due to the fact that the SHG measured in the experiment is the result of averaging over the laser spot (of about 50 $\mu$m in diameter) which contains hundreds of chaotically oriented crystallites each of them being several microns in size. Thus we can consider the symmetry of the graphene layer as that of an isotropic surface.

\begin{figure}[h]
\begin{center}
\includegraphics[width=7cm]{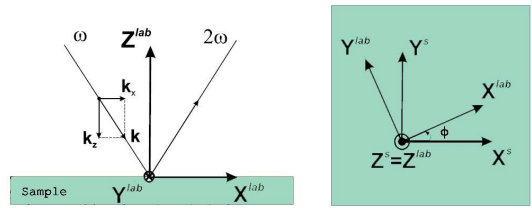}
\caption{Surface and laboratory coordinate systems}
\label{scheme}
\end{center}
\end{figure}

As the second step we will use the same symmetry of the CSHG and EFISH (electric field induced SHG) \cite{Vovochka_Eng} and this find nonzero $\chi^{(2,1)}$ tensor elements for the chosen experimental geometry. An important fact here is that these CSHG components are \textit{odd} in \textit{J}, i.e. change their sign under the reversal of the current direction to the opposite. Two coordinate system that are considered below are shown in Fig. \ref{scheme}, the laboratory one (XYZ) and that one connected with the sample, (X'Y'Z'). The axes Z and Z' are oriented along the normal to the surface while (XOY) and (X'OY') are parallel to the surface of the sample. Azimuthal rotation of the sample results in the modification of the angle $\phi$ between the X and X' axes. The plane of incidence corresponds to (XOZ), \textit{p} and \textit{s} polarizations of the light waves are also shown in the Figure.

In order to reveal the azimuthal CSHG dependence we first consider nonzero $\chi^{(2,1)}$ components for the DC current $\vec{J} \parallel$X',X ($\phi$=0), which are $\chi_{z'z'x'x'}, \chi_{z'x'z'x'}, \chi_{x'z'z'x'}, \chi_{x'x'x'x'}$ for the \textit{p}-in, \textit{p}-out polarizations. Let us consider the azimuthal rotation of the sample in that case. The SHG field in laboratory coordinate system generated by each of the $\chi^{(2,1)}$ nonzero components is determined by $E^{2\omega}_{i}=a_{ii'}a_{jj'}a_{kk'}a_{ll'}\cdot\chi^{(2,1)}_{i'j'k'l'}E^{\omega}_{j'}E^{\omega}_{k'}J_{l'}$, where $\hat{a}$ is a standard rotation matrix and the summation over the iterative indices is assumed. This gives us the following expression for the rotational anisotropy of the CSHG field  $E^{CSHG}_{pp} \propto cos\phi$. Analogously one can easily obtain $E^{CSHG}_{ps} \propto sin\phi$. 

Thus the azimuthal dependence for the CSHG contrast can be estimated. Taking into account that $\vec{E}^{2\omega}=\vec{E}^{2\omega}(\vec{J})+\vec{E}^{2\omega}(0)$ one can easily show that the CSHG contrast for the \textit{p}-in, \textit{p}-out combination of polarizations is determined by the ratio
\begin{eqnarray}\label{Psurf}
\rho_{pp,sp} = 2(I^{2\omega}(\vec J)-I^{2\omega}(0))/I^{2\omega}(0)\propto \\ {E}^{2\omega}(J)/E^{2\omega}(0) \propto cos\phi,\\
\rho_{ps,ss} \propto sin\phi,
\end{eqnarray}
In other words, maximal CSHG contrast in \textit{pp} and \textit{sp} combinations of polarizations should be observed as the DC current is flowing in the plane of incidence, $\vec{J}\parallel$X and should vanish in case $\vec{J}\parallel$Y. Analogously, for the case of \textit{s}-polarized SHG the CSHG contrast should be maximal for $\vec{J}\parallel$Y and vanish if  $\vec{J}\parallel$X. \\

\bibliography{Kursovaya}

\end{document}